\documentclass[aps,pre,twocolumn,superscriptaddress,showpacs,showkeys,floatfix]{revtex4}

\usepackage{amssymb}
\usepackage{amsmath}
\usepackage{bm}
\usepackage{graphicx}
\usepackage{grffile}
\usepackage{multirow}
\begin{document}
\setcounter{page}{1}
\title{Lattice Model for Spontaneous Imbibition in Porous Media:\\
The Role of Effective Tension and Universality Class}
\author{Deok-Sun \surname{Lee}}
\affiliation{Departments of Physics, Inha University, Incheon 402-751, Korea}
\author{Zeinab \surname{Sadjadi}}
\author{Heiko Rieger}
\affiliation{Theoretical Physics, Saarland University, 66041 Saarbr\"{u}cken, Germany}

\begin{abstract}
Recently anomalous scaling properties of front broadening during spontaneous imbibition of
water in Vycor glass, a nano-porous medium, were reported: the mean height and the width of the propagating front  increase with time $t$ both proportional to $t^{1/2}$.  
Here we propose a simple lattice imbibition model and elucidate quantitatively how the correlation range of the hydrostatic pressure and the disorder strength of the pore radii affect the scaling properties of the imbibition front. We introduce an effective tension of liquid across neighboring pores, which depends on the aspect ratio of each pore, and show that it leads to a dynamical crossover: both the mean height and the roughness grow faster 
in the presence of tension in the intermediate-time regime but eventually saturate in the long-time regime. The universality class of the long-time behavior is discussed by examining the associated scaling exponents and their relation to directed percolation.
\end{abstract}

\date{\today}
\maketitle 

\section{Introduction}

Moving interfaces in disordered media occur in various physical
situations and have been studied theoretically for some decades
now~\cite{barabasiBOOK95,meakinBOOK98}. It has been demonstrated 
that a few characteristics such as the embedding dimension, 
the conservation laws and the kind of non-linearity emerging
in a coarse grained equation of motion determine the scaling behavior
of the moving interfaces and establish a few universality classes
regardless of the microscopic details~\cite{barabasiBOOK95,meakinBOOK98,kpz86}.

Imbibition, the propagation of a fluid into the non-wetting region
found in, e.g., oil recovery, printing, irrigation, filtration,
etc~\cite{alava_imbibitionreview}, is also characterized by the motion
and morphology of the interface between the liquid and the non-wetting
region. Contrary to other interfaces, the theoretical understanding of
the imbibition front dynamics is far from complete, as it varies 
significantly between different experimental set-ups and between 
different models~\cite{rubio89,lenormand90,alava_imbibitionreview}. 
Moreover, the global conservation of the liquid volume generates
nontrivial correlations in the hydrostatic pressure in the liquid,
which affects the dynamics of the imbibition front. 

For spontaneous imbibition, when a wetting liquid is drawn into a
porous medium by capillary forces, it is well known that the balance
between the viscous drag and the pressure gradient in the bulk leads
to the Lucas-Washburn law~\cite{lucas18,washburn21} saying that the
average height of the imbibition front increases proportionally to the
square root of time. Far less clear is the roughening dynamics of the
imbibition front, i.e., the time dependence of the height fluctuations:
By focusing on the quenched disorder in the medium the relations to the
directed-percolation-depinning (DPD) model~\cite{tang92,buldyrev92},
the quenched Kardar-Parisi-Zhang universality (QKPZ) or the quenched
Edwards-Wilkinson (QEW) universality classes~\cite{leschhorn96} have
been suggested. But recently it was emphasized that in addition to
quenched disorder non-local interactions due to the fluid-conservation
law are crucial for the imbibition front roughening
\cite{lucas18,washburn21,krug91,aker98,ganesan98,dube99,chlam00,geromichalos02,gruener12}.
 
A novel universality class in the imbibition front broadening was recently
identified during the spontaneous imbibition of water in nanoporous
Vycor glass, a silica substrate of low porosity consisting of
nanometer-sized elongated pores~\cite{gruener12}: The mean height of 
the interface and its roughness are {\it both}
found to increase with time $t$ as $\sim t^{1/2}$. Since the ratio
of the roughness to the mean height usually decreases with time during
roughening dynamics, but is constant in this case, the interface
roughening appears anomalously strong. 

A pore-network model~\cite{gruener12}, in which liquid within the
pores propagates according to Hagen-Poiseuille's law, was capable to
reproduce the experimentally-observed features of the imbibition front
in Vycor glass. This suggested that the imbibition front consists of
disconnected menisci and that the hydrostatic pressure in the bulk and
the capillary pressure at these menisci are essential for
understanding the experimentally observed anomalous roughening. 
An essential feature of the pore-network model proposed in
\cite{gruener12} was the emergence of meniscus arrests at pore
junctions with branches of unequal radii. A subsequent scaling theory
for the meniscus arrest time distribution presented in
\cite{sadjadi13} predicted, in accordance with results from computer
simulations or the pore-network model, that arrest times of menisci in
the thicker branches of pore junction indeed scales with the height of
the junction where the arrest occurred. In this way the proposed theory 
could explain the proportionality between height and roughness
in random networks of elongated pores \cite{sadjadi13}.

Nevertheless the robustness of the observed scaling behavior is not
yet understood. Given many other kinds of kinetic interfaces
displaying scaling behavior different from those for imbibition, it
is natural to ask what factors are responsible for these
differences. To address this issue, we design in this paper a minimal
lattice model of imbibition and explore how the structure and dynamics
of the interface depend on the parameters. The proposed lattice
imbibition model has similarities with other lattice growth models but
involves as a new feature the hydrostatic pressure, which generates
long-range correlation. We adopt an approximation method of estimating
the hydrostatic pressure, which captures the essence of the pressure
profile generated by the liquid flow under the Hagen-Poiseuille's law
and the volume conservation.

We find that the disorder in the pore radii and the long correlation
length of the pressure, which will be explained in detail in the next
sections, are essential for the observed scaling behavior of
spontaneous imbibition. If the pore radii are uniform or the pressure
correlation spans only a short scale, the roughness becomes very weak
characterized by different scaling exponents from observed in the
experiment. It has been argued that the absence of an effective
tension acting across adjacent pores underlies the anomalous scaling
in the experiment. Computer simulations of our lattice imbibition
model shows that both the mean height and the roughness display a
crossover in time and eventually saturate when the effective tension
is present. In particular, they grow commonly as $t^{0.7}$ in the
intermediate-time regime before saturation, which is a faster growth with time
than without the tension. The formation of voids are suppressed due to
the effective tension, making the cluster of filled pores appear
compact. We discuss the intermediate-time behavior and the universality class
in connection with the properties of the directed percolation (DP) 
cluster.

The paper is organized as follows. The lattice imbibition model is
introduced in Sec.~\ref{sec:model}. The impact of the strength of the
pore radii and the lateral pressure correlation range is studied
in Secs.~\ref{sec:disorder} and \ref{sec:correlation}. In
Sec.~\ref{sec:tension} we perform the finite-size scaling analysis of
the mean height and the roughness as functions of time and present the
associated scaling exponents. Our findings are summarized and
discussed in Sec.~\ref{sec:discussion}.

\section{Model}
\label{sec:model}

\begin{figure}
\includegraphics[width=7cm]{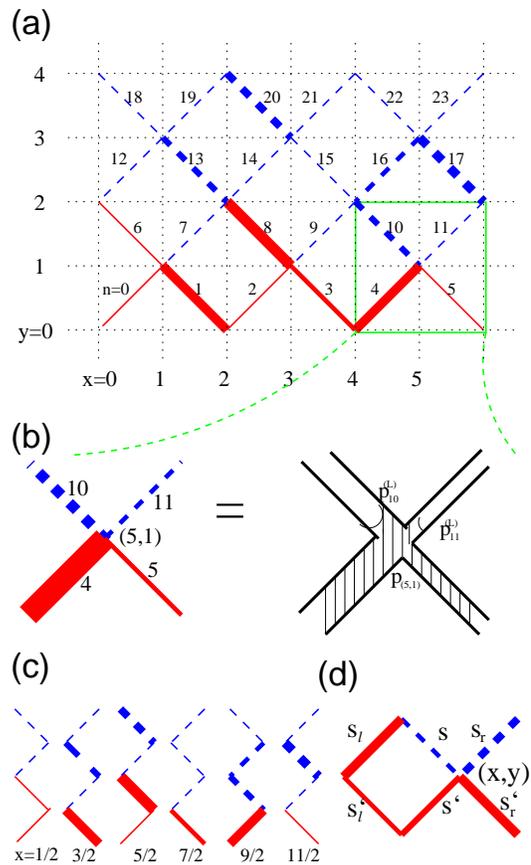}
\caption{Lattice imbibition model.  (a) Two-dimensional network of   $\mathcal{N}=N\times L$ pores with $N=6$ and $L=4$. Pores are indexed $n=0,1,2,\ldots, \mathcal{N}-1$ from the left bottom to the right top. Solid and dashed lines indicate filled and empty pores, respectively. 
The coordinates $(\bar{x}_n, \bar{y}_n)$ of the center of a pore $n$ are given by $\bar{x}_n = n - N \lfloor {n\over N}\rfloor +{1\over 2}$ and $\bar{y}_n = \lfloor {n\over N}\rfloor + {1 \over 2}$ and  with $\lfloor x\rfloor$ the integer not larger than $x$. The interface $\mathcal{I}$ consists of the pores $n=7,9,10,11,12,13,14$. (b)  $p^{\rm (L)}_{10}$ is the Laplace pressure at the free end of the meniscus in pore $n=10$ and 
$p_{(5,1)}$ is the hydrostatic pressure at $(5,1)$. (c) 6 independent vertical columns from (a). For the column at $x=5/2$, the pore $n=8$ is the highest filled pore, the upper end of which is at $y=2$, and  the pressure in the column is given by $p^{(0)}_{(5/2,y)}= (y/2)p^{\rm (L)}_8$ for $0\leq y< 2$ and $0$ for $y\geq 2$ according to Eq.~(\ref{eq:zerothorderpressure}). (d) The effective tension $T_{s,s'}=I$ in case of $f_{s_\ell}=1, f_{s_r}=0$, and $f_{s'_r}=1$.  }
\label{fig:model}
\end{figure}

We consider a two-dimensional network of  $\mathcal{N}=N \times L$ pores inclined at 45 degree as shown in Fig.~\ref{fig:model} (a).  
Each pore, indexed $n=0, 1, 2, \ldots, \mathcal{N}-1$ from the bottom left to the top right, has
radius $r_n$ selected randomly between $1-\Delta$ and $1+\Delta$ with $0\leq \Delta<1$. Here 
$\Delta$ characterizes the strength of the disorder in the pore radii. 
The boolean variable $f_n$ of pore $n$ takes $1$ or $0$ if it is filled or empty, respectively. 
The pressure $p_{(x,y)}$ represents the hydrostatic pressure at $(x,y)$.  Initially, all the pores are empty, $f_{n}=0$ for all $n$. The lower ends of the bottom pores $(0\leq n\leq N-1)$ are  immersed in the liquid such that the hydrostatic pressure at $y=0$ is kept constant, here taken as zero, i.e.,  $p_{(x,0)}=0$. The atmospheric pressure in the empty pores is also set to zero. Periodic boundary conditions are applied in $x$-direction.

The time-evolution of the liquid configuration $\{f_n\}$ describes the liquid propagation through the pore network. The interface $\mathcal{I}$ between wet and dry regions is identified with  the set of empty pores having at least one filled nearest-neighbor pore (see Fig.~\ref{fig:model} (a)). In order to update the liquid configuration $\{f_n\}$, an empty pore $s$ is chosen randomly among the interface pores. Then one of its filled neighbors $s'$ is selected. The driving force $P_{s,s'}$ acting on $s$ from $s'$ is computed. If $P_{s,s'}$ is positive, the pore $s$ is filled with probability $\max\{P_{s,s'},1\}$. If $P_{s,s'}$ is negative, the pore $s'$ is evacuated by the retraction of liquid with probability $\max\{|P_{s,s'}|,1\}$. 
This procedure is repeated as many as the number of interface pores, and then the macroscopic time $t$ is increased by $1$.

The liquid propagates spontaneously upward  (in $+y$ direction) due to the capillary pressure (see Fig.~\ref{fig:model} (a)). The pressure at the free end of the meniscus of pore $s$ is the Laplace pressure 
 \begin{equation}
p_{s}^{\rm (L)} = -{1\over r_{s}},
\label{eq:LaplacePressure}
\end{equation}
where $r_s$ is the radius of pore $s$ and we set the surface tension $\sigma=1/2$ in the general relation $p_s^{\rm (L)}=-2\sigma/r_s$. 
Note that here  the surface tension $\sigma$ is different from the effective tension acting across distinct pores considered in this work. 
Such low pressure at the menisci generates pressure gradient in the bulk and enables the liquid to propagate through the pores; $p_{(x,y)}$ decreases with $y$ ($p_{(x,0)}=0$). Let $(x,y)$ be the junction of the selected interface pore $s$ and its filled neighbor pore $s'$.  If liquid were to fill the pore $s$, the difference between the hydrostatic pressure at the junction and the Laplace pressure at the meniscus,
\begin{equation}
\Delta p_{s,s'} = \max\{p_{(x,y)} - p_s^{(L)},0\},
\label{eq:deltap}
\end{equation}
contributes to the driving force $P_{s,s'}$ pushing the liquid to fill the interface pore $s$ (See Fig.~\ref{fig:model} (b)).  If $p_{(x,y)}$ is lower than $p^{(L)}_s$, the pressure difference is taken to be zero as it does not contribute to the driving force~\cite{gruener12}.

\begin{figure}
\includegraphics[width=7cm]{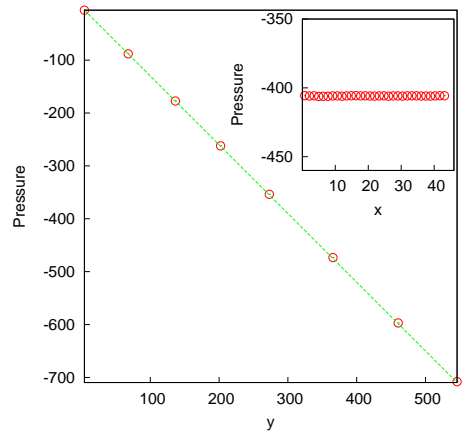}
\caption{Variation of the bulk pressure with position obtained from the pore-network model~\cite{gruener12, sadjadi13}. The pressure linearly decreases with height $y$ in a specific time. The dashed-line is a linear fit to the data. Inset: The pressure does not vary in $x$ direction at fixed $y$ (here $y/\langle h\rangle\thickapprox 1/3$).}
\label{fig:PXY}
\end{figure}

The pressure $p_{(x,y)}$  is determined by the boundary conditions,  liquid-volume conservation and Hagen-Poiseuille's law of incompressible fluid. Under the boundary conditions at the bottom ($p_{(x,0)}=0$) and the update rules described above, the pressure $p_{(x,y)}$ exhibits an important feature: it rarely varies with $x$ but decreases linearly with $y$ as illustrated in Fig.~\ref{fig:PXY} for the pore-network model in~\cite{gruener12, sadjadi13}. 

Instead of calculating the pressure field $p_{(x,y)}$ as in \cite{gruener12, sadjadi13} we use the following
simplified model for $p_{(x,y)}$. 
Let us first consider the simple case that all the vertical columns of pores are separated as in Fig.~\ref{fig:model} (c) and the pore $s$ is filled.
In each  column at $x'=1/2, 3/2, 5/2, \ldots, N-1/2$, the pressure $p^{(0)}_{(x',y)}$ decreases linearly from $0$ to $p^{(L)}_{\hat{n}_{x'}}$, where $\hat{n}_{x'}$ is the index of the highest filled pore in that column:
\begin{equation}
p_{(x',y)}^{(0)}=\left\{
\begin{array}{ll}
{y\over \hat{y}_{x'}} p^{\rm (L)}_{\hat{n}_{x'}} & {\rm for} \ y< \hat{y}_{x'},\\
0 & {\rm for} \ y \geq \hat{y}_{x'},
\end{array}
\right.
\label{eq:zerothorderpressure}
\end{equation}
where  $\hat{y}_{x'}$ is the $y$-coordinate of the upper end of the filled highest pore, indexed by $\hat{n}_{x'}$, in the given column at $x'$. In Eq.~(\ref{eq:zerothorderpressure}), we assumed that the pores beneath the highest filled one are filled. 

Since the columns at different lateral coordinates $x'$ are connected the pressure field 
$p_{(x,y)}$ will be laterally correlated due to fluid volume conservation. We therefore 
define the true hydrostatic pressure field $p_{(x,y)}$ to be an average over neighboring 
values of $p^{(0)}_{(x',y)}$'s at the same height: 
\begin{equation}
p_{(x,y)}={\sum_{|x-x'|<R}  p_{(x',y)}^{(0)}\over \sum_{|x-x'|<R} 1},
\label{eq:hydrostaticpressure}
\end{equation}
where $R$ is a model parameter showing the lateral correlation range of pressure. 
$R$ is expected to be large in the elongated-pore systems, as the pressure rarely varies with $x$.

Finally we define an effective tension as a force which reduces the energetic cost associated with the front width and depends on the aspect ratio of the pores. When the aspect ratio of each pore is small, the adjacent  menisci unite and form a continuous interface, thus the interface roughening is mainly slowed down by the effective surface tension. On the other hand, if the aspect ratio is very large, this effective tension is negligible as there is no continuous interface.
In our model, when the filled neighbor pore $s'$ is right below the interface pore $s$, the effective tension can affect the liquid propagation. In this case, the pore $s$ is more likely to be filled if its left or right neighbor is already filled. On the contrary,  if the left or right neighbor  of the filled pore $s'$ is empty, the pore $s'$ may be evacuated to reduce the interface length. These impacts of the effective tension can be modeled by  
\begin{equation}
T_{s,s'} = I \times  [ \max(f_{s_\ell},  f_{s'_\ell}-1) + \max (f_{s_r}, f_{s'_r}-1)], 
\label{eq:tension}
 \end{equation} 
where $s_\ell (s'_\ell)$  is the left neighbor of $s (s')$ and $s_r (s'_r)$ is the right neighbor of $s (s')$.  $T_{s,s'}$ takes a value among $-2I, -I,0, I$ and $2I$ depending on the liquid configuration around $s$ and $s'$. See Fig.~\ref{fig:model} (d) for an example. 

The driving force is the sum of the pressure difference, Eq.~(\ref{eq:deltap}), and the effective tension, Eq.~(\ref{eq:tension}):
\begin{equation}
P_{s,s'} = \Delta p_{s,s'} + T_{s,s'}.
\label{eq:Pss}
\end{equation}
The liquid fills $s$ with probability $\max\{P_{s,s'},1\}$ if $P_{s,s'}>0$  or is retracted from $s'$ with probability $\max\{|P_{s,s'}|,1\}$ if $P_{s,s'}<0$. 

\section{Disorder-induced strong roughening of the imbibition front}
\label{sec:disorder}
In this section, we study  the impact of the disorder in the pore radii on the scaling behavior of the mean height and the roughness. 
We perform simulations of the lattice imbibition model for different lateral system sizes $N$, pore radii disorder strengths $\Delta$, lateral correlation ranges of pressure $R$, and effective tension strengths $I$. The mean height $H$ and the interface width (roughness) $W$ are defined as 
\begin{align}
H &=\langle \overline{y}\rangle=\left\langle {1\over n(\mathcal{I})} \sum_{s\in \mathcal{I}} y_s\right\rangle,\nonumber\\
W &=\sqrt{\langle \overline{y^2} -\overline{y}^2 \rangle}\nonumber\\
&=\sqrt{\left\langle {1\over n(\mathcal{I})} \sum_{s\in \mathcal{I}} y_s^2 - \left({1\over n(\mathcal{I})} \sum_{s\in \mathcal{I}} y_s\right)^2\right\rangle},
\end{align}
where $y_s$ is the $y$-coordinate of the bottom end of a pore $s$, $\mathcal{I}$ is the interface and $n(\mathcal{I})$ is the number of the pores belonging to the interface. $\overline{\ldots}$ indicates the average over the interface pores and $\langle \ldots \rangle$ indicates the average over different realizations of simulation. We typically average over 1000 realizations.

Given that $p_{(x,0)}=0$ and ${p^{\rm (L)}}\sim -1$,  the pressure gradient $\Delta p_{s,s'}$ scales as  $1/H$. Without the effective tension, one finds that 
\begin{equation}
{dH\over dt} \sim \overline{\Delta p_{s,s'}}\sim {1\over H},
\end{equation}
which leads to the Lucas-Washburn's law $ H \sim \sqrt{t}$. In our simulations we observe the same behavior of height and pressure as above.
\begin{figure}
\includegraphics[width=7cm]{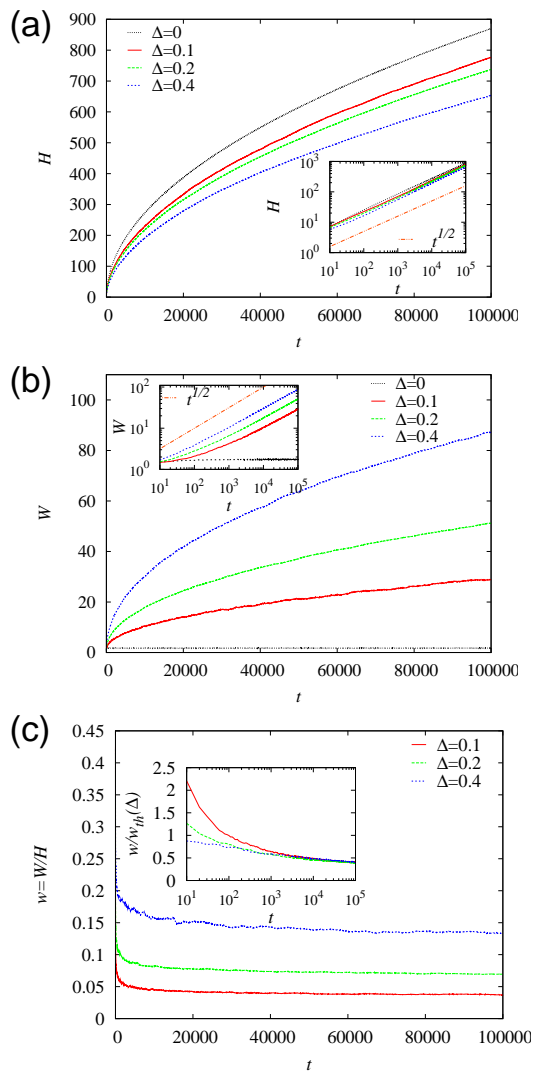}
\caption{Time-evolution of the mean height $H$ and the roughness $W$ for $N=32, R=N/2, I=0$ and different values of $\Delta$. (a) Plots of $H$ versus time $t$. Inset: The same plots in logarithmic scales showing that $H\sim t^{1/2}$. The dashed line has slope $1/2$. (b)  Plots of $W$ versus $t$. Inset: The same plots in logarithmic scales showing that $W\sim t^{1/2}$. The smaller $\Delta$ is, the later the regime showing the square-root scaling appears. The dashed line has slope $1/2$. (c) Plots of the relative roughness $w={W \over H}$ versus $t$. Inset: The ratio ${w\over w_{\rm th}(\Delta)}$ becomes independent of $\Delta$ in the long-time limit as predicted in Eq.~(\ref{eq:wth2}).}
\label{fig:HWI0}
\end{figure}
We set $R=N/2$, i.e., the lateral correlation range of pressure equals the system size, as the sum in Eq.~(\ref{eq:hydrostaticpressure}) runs over all possible $x'$ values. We assume $I=0$, i.e., there is no effective tension.
 Simulation results for average height and roughness are plotted in Figs.~\ref{fig:HWI0} (a) and (b), respectively. We find that
\begin{equation}
H\sim t^{B}, \ \ B =0.53 (5)
\label{eq:Ht0}
\end{equation}
and 
\begin{equation}
 W\sim t^{\beta}, \ \ \beta= 0.46 (8).
\label{eq:Wt0}
\end{equation}

\begin{figure}
\includegraphics[width=5cm]{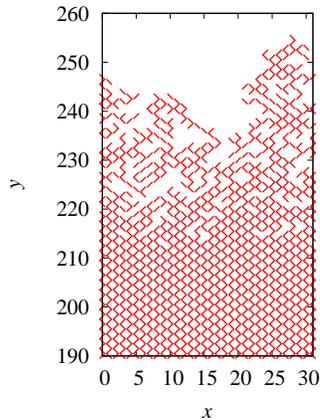}
\caption{Liquid configuration at time $t=10^4$ in the lattice imbibition model for $N=32, \Delta=0.1, R=N/2$, and $I=0$.}
\label{fig:config}
\end{figure}

The values of the scaling exponents $B$ and $\beta$ are both close to  $1/2$, in agreement with the results obtained for the pore-network model in ~\cite{gruener12}. The roughness $W$ and the ratio $W/H$ increase with disorder strength $\Delta$,  while the average height $H$ decreases in accordance with ~\cite{gruener12}. For an ordered  lattice i.e., \ $\Delta=0$, the roughness remains at a small value, meaning that the interface is smooth, while the mean height increases. A typical liquid configuration for a disordered lattice  is shown in  
Fig.~\ref{fig:config}. One observes a large number voids generated by thick pores with a Laplace pressure $p^{\rm (L)}$ that is larger than the  hydrostatic pressure at the junction, preventing the filling of the pore~\cite{sadjadi13}. For any non-vanishing disorder strength $\Delta>0$ the interface is rough and the ratio $W/H$ is non-zero.

Along the boundaries of voids the Laplace pressure $p^{\rm (L)}$ is larger than the 
hydrostatic pressure within the nearest pore junction. Since the hydrostatic pressure $p_{(x,y)}$ decreases with increasing $y$, the voids are more likely to emerge as the height $y$ increases. The lowest possible height, $y_{\rm min}$,  for which voids may exist can be obtained by equating the hydrostatic pressure and the largest possible value of the Laplace pressure $p^{\rm (L)}_{\rm (max)}$. The hydrostatic pressure at $y$ is estimated as 
 $\langle p\rangle(y)  \simeq  \langle p^{(L)}\rangle {y\over H}$ with $H$ the mean height. Therefore at $y=y_{\rm min}$, it holds that 
 \begin{equation}
 \langle p^{\rm (L)}\rangle {y_{\rm min} \over H} = p^{\rm (L)}_{\rm max}.
\label{eq:ymin}
 \end{equation}
The pore radii are distributed uniformly between $1-\Delta$ and $1+\Delta$, and therefore $\langle p^{\rm (L)}\rangle = - \langle {1\over r}\rangle = - {1\over 2\Delta} \ln {1+\Delta\over 1-\Delta}$ and $p_{\rm max}^{\rm (L)} = -{1\over 1+\Delta}$. Using these relations in Eq.~(\ref{eq:ymin}), we find that ${H\over y_{\rm min}} = { 1+\Delta\over 2\Delta }\ln ({1+\Delta \over 1-\Delta})$. $H-y_{\rm min}$ is an estimate for
the front width and the relative width $w_{\rm th}$ obeys
\begin{align}
w_{\rm th} & \simeq  {H-y_{\rm min} \over H} = 1 - {2\Delta \over 1+\Delta}\left[ \ln \left({1+\Delta\over 1-\Delta}\right)\right]^{-1}.
\label{eq:wth}
\end{align}
We conjecture that the relative roughness $w=W/H$ scales in the same way as $w_{\rm th}$
\begin{equation}
w\simeq w_{\rm th}\simeq\Delta\quad{\rm for}\quad\Delta\ll 1,
\label{eq:wth2}
\end{equation}
which is consistent with the simulation results in Fig.~\ref{fig:HWI0} (b) and (c), where $W/H \simeq 0$ for $\Delta=0$ and $W/H$ increases with $\Delta$. In addition the long-time limit of $w/w_{\rm th}$ becomes independent of $\Delta$ as shown in the inset of Fig.~\ref{fig:HWI0} (c) in agreement with Eq.~(\ref{eq:wth2}). From the simulation results and the analytical argument we conclude that the disorder in the pore radii is crucial for the interface roughening of imbibition in the elongated-pore systems.

\section{Impact of the pressure correlation}
\label{sec:correlation}

\begin{figure}
\includegraphics[width=7cm]{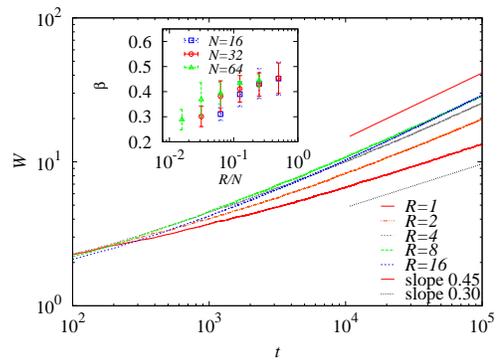}
\caption{Roughness $W$  as a function of time $t$ for $N=32, I=0, \Delta=0.1$ and different ranges $R$ of pressure correlation. The roughness increases with time as $W\sim t^\beta$ with $\beta$ increasing with $R$. The solid and dotted lines have slope 0.45 and 0.3, respectively. Inset: The exponent $\beta$ as a function of $R/N$ for different $N$'s.}
\label{fig:WforDifferentR}
\end{figure}

In this section, we explore the impact of the correlation range of pressure $R$ on the roughness of imbibition front.
If $R=1$,  only the zeroth-order pressures $p^{(0)}$ of the two pores touching the junction point $(x,y)$ are used to evaluate $p_{(x,y)}$ according to Eq.~(\ref{eq:hydrostaticpressure}). $p_{(x,y)}$ may fluctuate not only with $y$ but also with $x$ as if the vertical columns were separated. On the contrary, if $R\geq N/2$, $p_{(x,y)}$ is the average of $p^{(0)}_{(x',y)}$'s for all $x'$, which does not vary with $x$ for given $y$. 
It seems plausible to assume that the liquid-volume conservation in the elongated-pore systems makes the lateral correlation of pressure long-ranged, i.e.\ $R\geq N/2$ in our model. 

For $R=1$ one expects a roughening exponent $\beta=1/4$: In this case the height in each column
evolves with time independently from the neighbor and one can expect random deposition behavior, in which
case the height-height fluctuations between the columns evolve as $W\sim\sqrt{\langle h\rangle}\sim t^{1/4}$,
which can also be seen as follows. 
The interface height $h$ in a single column evolves with time as ${d h\over dt} = {1\over\xi(h) h}$, where 
$\langle \xi\rangle = c$ and $\langle \xi(h)\xi(h')\rangle - \langle \xi\rangle^2 = d \delta(h-h')$ with $c$ and $d$ some constants~\cite{gruener12}. The time taken for the interface to reach $h$ satisfies statistically $ \langle t\rangle \sim h^2$ and $\langle t^2\rangle -\langle t\rangle^2\sim   h^{3/2}$, which yields equivalently $\langle h\rangle \sim t^{1/2}$ and $W\sim \sqrt{\langle h^2\rangle - \langle h\rangle^2} \sim t^{1/4}$~\cite{gruener12}. 

Our lattice imbibition model demonstrates that the mean height $H(t)$ is not affected by the correlation range $R$. On the other hand, the roughness is strongly affected by $R$. The scaling exponent $\beta$ introduced in Eq.~(\ref{eq:Wt0}) turns out to be very close to  $1/4$ for $R=1$ and increases to $1/2$ with increasing $R$ as shown in Fig.~\ref{fig:WforDifferentR}. The data shown Fig.~\ref{fig:WforDifferentR} indicate that in the
infinite system size limit $N\to\infty$ the universality class of the roughening process changes from the random deposition universality class ($\beta=1/4$) for finite $R$ (i.e. $R/N\to0$) to the spontaneous imbibition universality class ($\beta=1/2$) for infinite $R$ (i.e., $R\propto N)$.

\section{Impact of the effective tension}
\label{sec:tension}

\begin{figure}
\includegraphics[width=7cm]{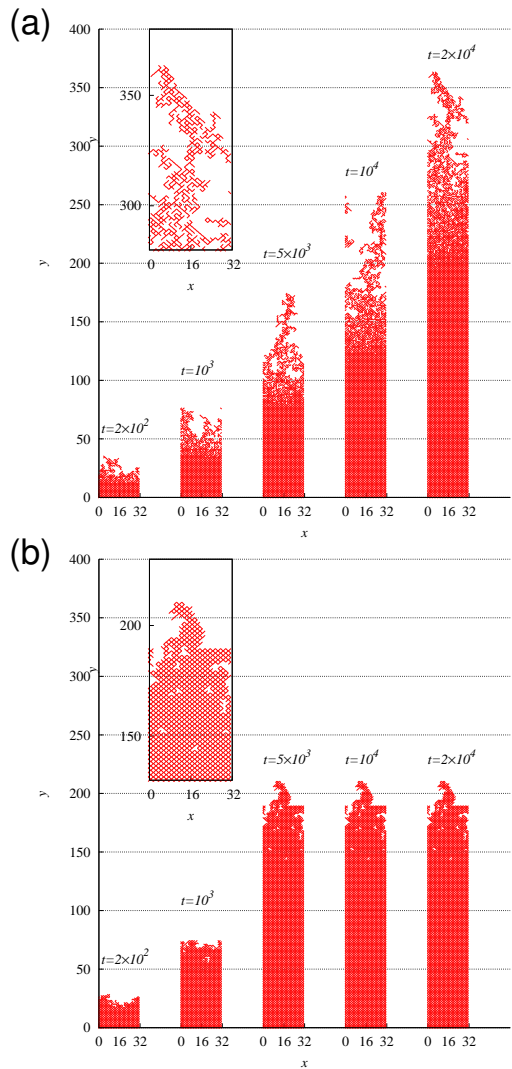}
\caption{Time evolution of the Êliquid configuration with (a) $I=0$ and (b) $I=0.08$ for $N=32, \Delta=0.4, R=N/2$. The insets show magnified views at $t=2\times 10^4$ (a) for $280\leq y\leq 380$ and (b) for $130\leq y\leq 230$.  The effective tension makes the cluster of filled pores more compact than without tension and leads the front to stop in the long-time limit. }
\label{fig:config2}
\end{figure}

\begin{figure}
\includegraphics[width=8cm]{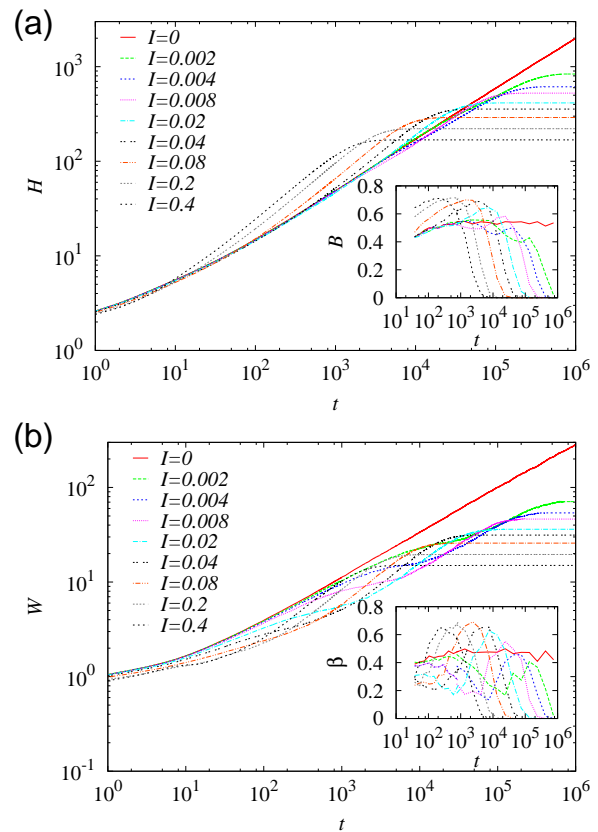}
\caption{Time-evolution of the mean height $H$ and the roughness $W$ under the effective tension for $N=64, R=N/2, \Delta=0.4$ and different values of $I$. (a) Plots of $H$ versus time $t$. 
$H$ grows faster with $I>0$ than with $I=0$ in the intermediate-time regime and saturates in the long-time limit. Inset: The estimated exponent $B$ in the relation $H\sim t^B$  increases from 0.5 to 0.7 before the saturation of $H$. (b) Plots of $W$ versus $t$. The roughness is not larger with $I>0$ than that with $I=0$ in the whole time regime. As the mean height does, the roughness increases faster with $I>0$ than with $I=0$ in the intermediate-time regime and saturates in the long-time limit. Compared with the mean height, the time-dependent behavior of $W$ are complicated.  Inset: The estimated exponent $\beta$ in the relation $W\sim t^\beta$ finds its maximum value around 0.7. }
\label{fig:HWIall}
\end{figure}

The driving force $P_{s,s'}$ consist of the effective tension $T_{s,s'}$ as well as the pressure difference $\Delta p_{s,s'}$. $T_{s,s'}$  in Eq.~(\ref{eq:tension}) ranges between $-2I$ and $2I$, thus, it may contribute to both filling an empty pore $s$ or emptying a filled pore $s'$. While the effective tension  remains of order $I$, the pressure difference decreases with time as $\Delta p_{s,s'}\sim H^{-1}$.  The evolution of the liquid configurations with and without the effective tension at different times are presented in Fig.~\ref{fig:config2} . The time-evolution of the mean height $H$ and the roughness $W$ are shown  in Fig.~\ref{fig:HWIall} for different strengths of the effective tension $I$.  The most important impact of tension is that the imbibition front propagation stops at some point and its roughness saturates.  Moreover, the tension smoothes the interface. The roughness is smaller with tension ($I>0$) than without it ($I=0$). The clusters of filled pores in presence of tension are more compact than those without tension (see Fig.~\ref{fig:config2}). For $I\ll 1$, in the early-time regime the pressure differences are larger than the effective tension and the dynamical evolution of height and roughness is similar to the case without tension ($I=0$). For longer times the pressure differences decrease and become comparable to the
effective tension. In the intermediate-time regime, $H$ and $W$ grow faster with tension than without. The liquid propagation eventually stops because the pressure differences drop below the effective tension. The larger $I$, the earlier the long-time regime begins and the smaller the saturation values of $H$ and $W$.  These observations also depend on the system size $N$.

To understand the impact of the effective tension on the  roughening of the imbibition front we explore quantitatively the behavior of  $H$ and $W$ on $I$ and $N$. We take the scaling ansatz for the behavior of $H$ and $W$ and estimate the associated scaling exponents by their data collapse around the boundaries of the distinct time regimes, which allows us to see the width of each time regime.  As the mean height and the roughness behave in different ways, we study them separately. 

\subsection{Scaling behavior of the mean height $H$} 
As shown in Fig.~\ref{fig:HWIall} (a), three regimes appear in the time evolution of $H$ with characteristic time scales $t_1$ and $t_H$.  In the short-time regime, $t\ll t_1$, $H \sim t^{B}$ with the same exponent as in Eq.~(\ref{eq:Ht0}) regardless of $I$. In the intermediate-time regime,  $t_1\ll t\ll t_H$, the mean height grows faster than the former case, i.e., $H \sim t^{B_{*}}$ with $B_{*}>B$.  In the long-time limit,  $t\gg t_H$, the liquid does not propagate anymore and the height saturates to $H_{\rm sat}$. In summary
\begin{align}
H &\sim \left\{
\begin{array}{ll}
t^B & \ {\rm for} \ t\ll t_1, \\
t^{B_*} & \ {\rm for} \ t_1\ll t\ll t_H,\\
H_{\rm sat} & \ {\rm for} \ t\gg t_H.
\end{array}
\right.
\label{eq:Hall}
\end{align} 

The effective scaling exponent $B(t)$, given in the inset of Fig.~\ref{fig:HWIall} (a),  can be  estimated as $B(t) = \ln [{H(t+\delta t)\over H(t)}]/\ln [{t+\delta t\over t}]$ for a time window $\delta t$ around time $t$. By averaging  $B(t)$ around its maximum,  we  estimate $B_*$ as~\footnote{$B$ is estimated by the average value of $B(t)$ in the very early-time regime and $B_{*}$ by the average value of $B(t)$ around the time when $B(t)$ is maximum. These average values still vary with $I$ and $N$, and these variations are included in their errorbars.}
\begin{equation}
B_*=0.70(5),
\label{eq:B*}
\end{equation}
for $\Delta=0.4$, which is not significantly different from $B_*=0.73(10)$ for $\Delta=0.1$. 

\begin{table}
\begin{tabular}{c|c|c||c|c|c}
&\multicolumn{2}{c||}{$H$}&&\multicolumn{2}{|c}{$W$}\\
\hline
                 & $\Delta=0.1$     & $\Delta=0.4$ && $\Delta=0.1$     & $\Delta=0.4$\\
                 \hline
$B$           &$0.53(5)$          &$0.53(3)$  & $\beta$    &$0.46(5)$ &$0.46(6)$\\
$B_*$        &$0.73(10)$        &$0.70(5)$  & $\beta_*$ &$0.76(10)$ &$0.67(5)$\\
$\eta_1$    &$2.0(1)$           &$2.0(1)$    & $\eta_2$  &$1.2(1)$ &$1.2(1)$\\
$\alpha_H$ &$1.0(1)$           &$0.4(1)$    & $\alpha_W$ &$1.1(1)$ &$0.5(1)$ \\
$\nu_H$     &$0.47(10)$       &$0.36(4)$  & $\nu_W$ &$0.45(9)$ &$0.33(8)$\\
$\eta_H$    &$1.2$                &$1.0$         & $\eta_W$ &$1.3$ &$1.1$\\
$z_H$         &$1.4$               &$0.6$         & $z_W$ &$1.1$ &$0.45$ \\ 
\multicolumn{3}{c||}{}                                                            & $\theta$ &$0.40(5)$ &$0.43(5)$ \\
\multicolumn{3}{c||}{}   								& $\zeta$ &$0.3(1)$ &$0.2(1)$
\end{tabular}
\caption{The scaling exponents for the mean height $H$ and the roughness $W$. See the main text for their definitions.}
\label{table:exponents}
\end{table}

\begin{figure}
\includegraphics[width=7cm]{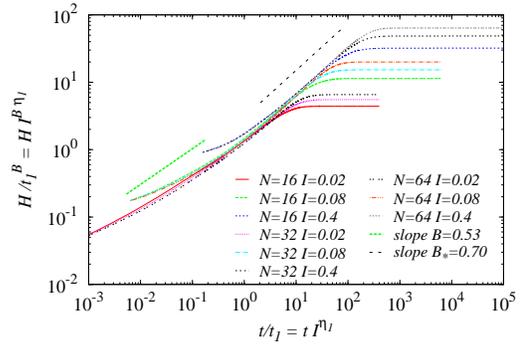}
\caption{Collapse of the scaled data $H/t_1^B = HI^{B\eta_1}$ as functions of  $t/t_1=tI^{\eta_1}$ with $\eta_1=2.0$ and $B=0.53$ for  $\Delta=0.4$ and different values of system size $N$ and the effective tension $I$. See Eq.~(\ref{eq:H1collapse}). Two lines have slopes $B=0.53$ and $B_*=0.70$, respectively, as in Eqs.~(\ref{eq:Ht0}) and (\ref{eq:B*}).}
\label{fig:Hscal1}
\end{figure}

We find that around the first crossover time $t_1$, the simulation data for different values of $I$ and $N$ collapse onto a single curve showing the crossover behavior under the following scaling form:
\begin{align}
H &=  t_1^B \Phi_1 \left({t\over t_1}\right), \ {\rm where} \nonumber\\
t_1 & = I^{-\eta_1} \ {\rm and} \nonumber\\
\Phi_1 (x) &\sim \left\{
\begin{array}{ll}
x^B & \ {\rm for } \ x\ll 1,\\
x^{B_*} & \ {\rm for} \ x\gg 1. 
\end{array}
\right. 
\label{eq:H1collapse}
\end{align}
The corresponding scaled data are shown in  Fig.~\ref{fig:Hscal1}. The exponent $\eta_1$ equals $2.0(1)$. 

\begin{figure}
\includegraphics[width=7cm]{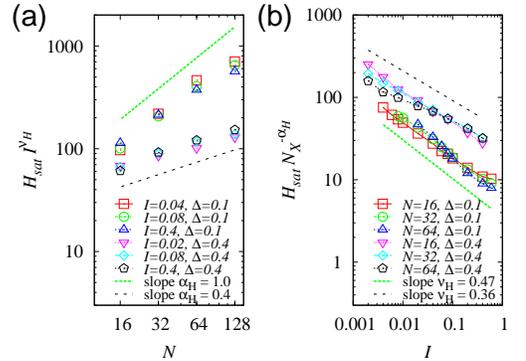}
\caption{Scaling behavior of the saturation height $H_{\rm sat}$ with respect to the system size $N$ and the effective tension strength $I$.  
(a) Collapse  of the plots of $H_{\rm sat} I^{\nu_H}$ versus $N$ for different $I$'s and given $\Delta$. The upper collapsed data are for $\Delta=0.1$ and the lower for $\Delta=0.4$.  (b) Collapse of $H_{\rm sat} N^{-\alpha_H}$ versus $I$ for different $N$'s and given $\Delta$. The upper collapsed data are for $\Delta=0.4$ and the lower for $\Delta=0.1$. The data collapse in  both plots are found with $\nu_H=0.47$ and $\alpha_H = 1.0$ for $\Delta=0.1$ and $\nu_H=0.36$ and $\alpha_H = 0.4$ for $\Delta=0.4$.}
\label{fig:Hinf}
\end{figure}

After the fast growth in the intermediate regime ($H\sim t^{B_*}$), $H$ eventually saturates to its maximum value over the crossover time $t_H$. The saturation values shown in Fig.~\ref{fig:Hinf} scales with the system size $N$ and the effective tension $I$ as 
\begin{equation}
H_{\rm sat} \sim I^{-\nu_H} N^{\alpha_H}.
\label{eq:Hsat}
\end{equation}
The values of exponents $\nu_H$ and $\alpha_H$ are given in Table~\ref{table:exponents} for $\Delta=0.1$ and $0.4$.

\begin{figure}
\includegraphics[width=7cm]{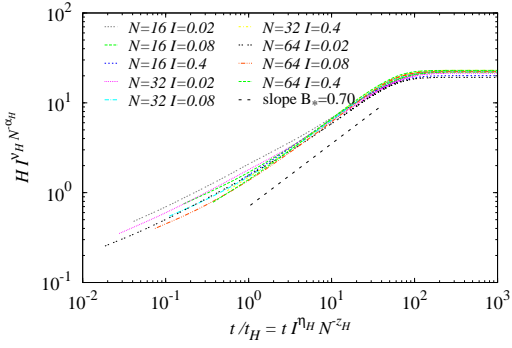}
\caption{Collapse of the scaled data $H I^{\nu_H}N^{-\alpha_H}$ as functions of the scaling variable $t/t_H=t I^{\eta_H} N^{-z_H}$ for $\Delta=0.4$ and different values of $N$ and $I$. See Eq.~(\ref{eq:HHcollapse}). The values of the scaling exponents are $\nu_H=0.36, \alpha_H = 0.4, \eta_H = 1.0$ and $z_H = 0.6$. The dashed  line has slope $B_*=0.70$.}
\label{fig:Hscal2}
\end{figure}

Using Eq.~(\ref{eq:Hsat}), one finds the following scaling functions for $H(t)$ in the  case of finite $t/t_H$:
\begin{align}
H &= I^{-\nu_H} N^{\alpha_H} \Phi_H \left({t\over t_H}\right),\ {\rm where} \nonumber\\
t_H & = I^{-\eta_H} N^{z_H} \ {\rm and} \nonumber\\
\Phi_H (x) &\sim \left\{
\begin{array}{ll}
x^{B_*} & \ {\rm for } \ x\ll 1,\\
1 & \ {\rm for} \ x\gg 1
\end{array}
\right.  \ {\rm with} \nonumber\\
\eta_H &= {\eta_1 (B_*-B) +\nu_H \over B_*} \ {\rm and} \nonumber\\
z_H &= {\alpha_H\over B_*}.
\label{eq:HHcollapse}
\end{align}
The data collapse  are shown in Fig.~\ref{fig:Hscal2}. All exponents obtained from the simulation are presented in Table~\ref{table:exponents}.

Equations~(\ref{eq:H1collapse}) and (\ref{eq:HHcollapse}) indicate that the fast growth  $H\sim t^{B_*}$ with $B_*\sim 0.7$ appears in a wide range of times when $I$ has a non-zero finite value and $N$ is large: the intermediate regime begins at $t_1=I^{-\eta_1}=\mathcal{O}(1)$ and ends at $t_H = I^{-\eta_H} N^{z_H}\gg 1$ for $N\gg 1$.  The intermediate  regime is observed as long as $t_H\gg t_1$ or equivalently  
\begin{equation}
I \gg  N^{- {z_H\over \eta_1 - \eta_H}}.
\label{eq:Hcond}
\end{equation}
If the effective tension is so small as to violate Eq.~(\ref{eq:Hcond}), the square-root growth of the mean height would persist until saturation occurs.

\subsection{Scaling behavior of the roughness $W$} 

The behavior of the roughness is more complicated than the mean height as seen in Fig.~\ref{fig:HWIall} (b). Here, four regimes are observed which display distinct behavior of $W$ as
\begin{align}
W &\sim \left\{
\begin{array}{ll}
t^\beta & \ {\rm for} \ t\ll t_0, \\
{\rm slowly} \ {\rm increasing} & \ {\rm for} \ t_0\ll t\ll t_2,\\
t^{\beta_*} & \ {\rm for} \ t_2\ll t\ll t_W,\\
W_{\rm sat} & \ {\rm for} \ t\gg t_W.
\end{array}
\right.
\label{eq:Wall}
\end{align} 
Among these regimes, the first two are rather difficult to identify.  We expect  $W\sim t^\beta$ in the first regime, $t\ll t_0$, but its duration is rather short.
For the same reason, the behavior of $W$  is not clearly seen in the second regime. To make these early-time regimes long enough, the model with much smaller values of $I$ for much longer time  should be simulated than in this study. We restrict ourselves to the scaling behavior of $W$ around $t=t_2$ and $t=t_W$.  In the third regime, $t_2\ll t\ll t_W$, fast growth $W\sim t^{\beta_*}$ with $\beta_*> \beta \sim 0.5$ is observed, which eventually saturates over $t\gg t_W$.

We estimate the effective scaling exponent $\beta(t)$,  (inset of Fig.~\ref{fig:HWIall} (b)), in a similar way we argued for height. While we observe a nearly constant value $\beta(t)\simeq 0.46$ for $I=0$ for the whole time period, $\beta(t)$ varies with time when $I>0$ . We find $\beta_* = 0.67(5)$ for $\Delta=0.4$ and $\beta_*=0.76(10)$ for $\Delta=0.1$.

\begin{figure}
\includegraphics[width=7cm]{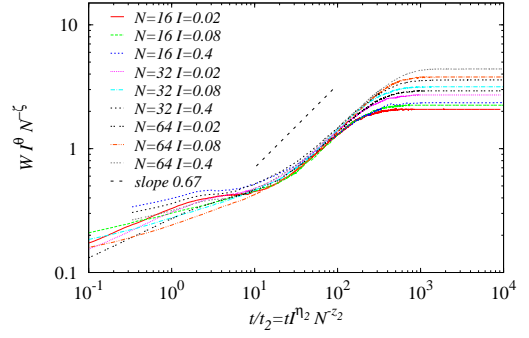}
\caption{Collapse of the scaled data of $W I^\theta N^{-\zeta}$ as functions of  $t/t_2=tI^{\eta_2}$ for $t/t_2\gg 1$ with $\eta_2=1.2$, $\theta=0.43,$ and $\zeta=0.2$ for  $\Delta=0.4$ and different values of system size $N$ and the effective tension $I$. See Eq.~(\ref{eq:W1collapse}). The line has slope $\beta_*=0.67$ as in Eq.~(\ref{eq:Wt0}) .}
\label{fig:Wscal1}
\end{figure}

Around $t=t_2$, the curvature of $W(t)$ changes and $W(t)$ grows as $t^{\beta_*}$ for $t\gg t_2$. Assuming that the value of $W(t_2)$ scales with the tension $I$ and the system size $N$ as $W(t_2) \sim I^{-\theta} N^\zeta$, where $\theta$ and $\zeta$ are new scaling exponents, we find that the scaled data of the roughness for different $N$'s and $I$'s collapse onto a curve in the regime $t\gg t_2$ as
\begin{align}
W &=  I^{-\theta} N^\zeta \ \Phi_2 \left({t\over t_2}\right), \ {\rm where} \nonumber\\
t_2 & = I^{-\eta_2}, \ {\rm and} \nonumber\\
\Phi_2 (x) &\sim x^{\beta_*} \ {\rm for} \ x\gg 1
\label{eq:W1collapse}
\end{align}
as presented in  Fig.~\ref{fig:Wscal1}. Here the scaling exponents are $\theta\simeq 0.43, \zeta\simeq 0.2$, and $\eta_2\simeq 1.2$ for $\Delta=0.4$. In these exponents there is no significant difference  between  $\Delta=0.1$ and $\Delta=0.4$ (See Table~\ref{table:exponents}).

\begin{figure}
\includegraphics[width=8cm]{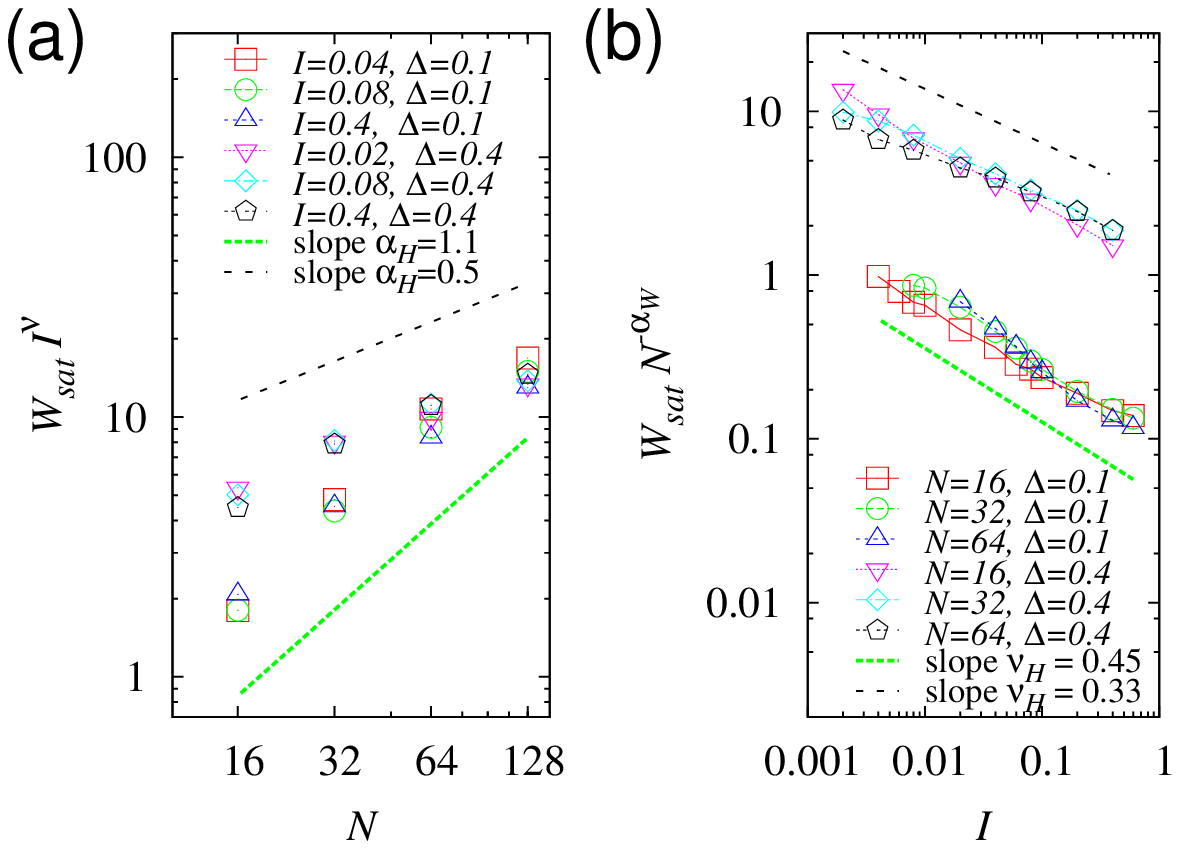}
\caption{
Scaling behavior of the saturation roughness $W_{\rm sat}$ with respect to the system size $N$ and the effective tension strength $I$.  
(a) Collapse  of the plots of $W_{\rm sat} I^{\nu_W}$ versus $N$ for different $I$'s and given $\Delta$. The upper collapsed data are for $\Delta=0.4$ and the lower for $\Delta=0.1$.  (b) Collapse of $W_{\rm sat} N^{-\alpha_W}$ versus $I$ for different $N$'s and given $\Delta$. The upper collapsed data are for $\Delta=0.4$ and the lower for $\Delta=0.1$. The data collapse in  both plots are found with $\nu_W=0.45$ and $\alpha_W = 1.1$ for $\Delta=0.1$ and $\nu_W=0.33$ and $\alpha_W = 0.5$ for $\Delta=0.4$.}
\label{fig:Winf}
\end{figure}

\begin{figure}
\includegraphics[width=7cm]{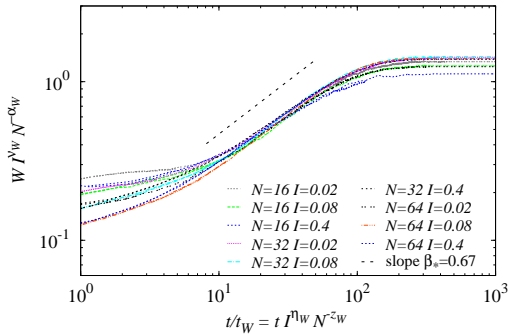}
\caption{
Collapse of the scaled data $W I^{\nu_W}N^{-\alpha_W}$ as functions of the scaling variable $t/t_W=t I^{\eta_W} N^{-z_W}$ for $\Delta=0.4$ and different values of $N$ and $I$. See Eq.~(\ref{eq:WWcollapse}). The scaling exponents are $\nu_W=0.33, \alpha_W = 0.5, \eta_W = 1.1$ and $z_W = 0.45$. The dashed  line has slope $\beta_*=0.67$.
The collapse of the scaled data $W I^{\nu_W}N^{-\alpha_W}$ as functions of the scaling variable $t/t_W=t I^{\eta_W} N^{-z_W}$ for different values of $N$ and $I$ as given in  Eq.~(\ref{eq:WWcollapse}). Here, for $\Delta=0.4$,  we used $\eta_W=1.1$ and $z_H = 0.45$. The dashed  line has slope $\beta_*=0.67$.}
\label{fig:Wscal2}
\end{figure}

The saturation value of the roughness $W_{\rm sat}$  for $t\gg t_W$  scales as 
\begin{equation}
W_{\rm sat} \sim I^{-\nu_W} N^{\alpha_W}
\label{eq:Wsat}
\end{equation}
with the exponents given by $\nu_W =0.33(8)$ and $\alpha_W = 0.5(1)$ for $\Delta=0.4$ as shown in Fig.~\ref{fig:Winf}. 
Around $t_W$, the following scaling function $\Phi_W(x)$ characterizes the behavior of $W$:
\begin{align}
W &= I^{-\nu_W} N^{\alpha_W} \Phi_W \left({t\over t_W}\right), \ {\rm where} \nonumber\\
t_W & = I^{-\eta_W} N^{z_W} \ {\rm and}   \nonumber\\
\Phi_W(x) &\sim \left\{
\begin{array}{ll}
x^{\beta_*} & \ {\rm for } \ x\ll 1,\\
1 & \ {\rm for} \ x\gg 1
\end{array}
\right. \ {\rm with} \nonumber \\
\eta_W &= \eta_2 -{\theta -\nu_W \over \beta_*}\ {\rm and}\nonumber\\
z_W &= {\alpha_W-\zeta \over \beta_*}.
\label{eq:WWcollapse}
\end{align}
The resulting scaled data are presented in Fig.~\ref{fig:Wscal2} and the exponents are given in Table~\ref{table:exponents}.

The  mean height and the roughness have quite similar scaling properties in the intermediate- and long-time regimes;
both $H$ and $W$ grow as a power law with $B_*,\beta_*\simeq 0.7$ which is larger than $0.5$ obtained in absence of effective tension $I$, and saturate at the values which scale as $I^{-\nu} N^\alpha$  with $\nu\simeq 1$ and $0.4\lesssim \alpha\lesssim 1.4$ and thus diverge in the limits $N\to \infty$ and $I\to 0$.  

\subsection{Understanding the scaling behavior in the intermediate and long times}

The scaling of interface height and roughness in the intermediate-time regime, their saturation in the long-time regime, as well as the values of the exponents $B_*$ and $\beta_*$ ($\simeq 0.7$),  are reminiscent of what is observed during the motion of interfaces in random environments close to the depinning transition point~\cite{tang92}. The dynamics and roughening of the interface vary with the density of pinning centers and  are governed by the mechanism of directed percolation (DP) in the vicinity of the pinning-depinning transition point. The pinned nodes form a cluster which is qualitatively equivalent to the DP cluster, and the scaling behavior of the latter determine various properties of the interface. In the imbibition model studied in this work, the driving force is provided by the Laplace pressure,  which depends on the pore radii and is thus a random force. Pinning appears when the hydrostatic pressure is much lower than the Laplace pressure of the meniscus which is relevant in thick pores, and when the effective tension forbids the liquid from propagating or even retracts the liquid that has proceeded.

Let us compare our results to the scaling properties of the DP clusters. Suppose the pinning cluster, the cluster of pores suppressing the liquid propagation, is as large as $\xi_\perp$ in the perpendicular ($y$) direction and as $\xi_\parallel$ in the lateral ($x$) direction. If this cluster is a DP cluster, those length scales scale as
\begin{equation}
\xi_\perp \sim |q-q_c|^{-\nu_\perp}, \ \ \xi_\parallel\sim |q-q_c|^{-\nu_\parallel}
\end{equation}
with $q$ being the density of pinning center and $q_c$ the critical point. The mean height and the roughness  are set commonly by $\xi_\perp$ and the time scale is set by $\xi_\parallel$ as $H\sim W\sim \xi_\perp \Phi(t/\xi_\parallel)$~\cite{tang92}. If the pinning density is so close to the critical point that $\xi$ is larger than the lateral system size $N$, the lateral size of the cluster is set by $N$ and 
\begin{align}
H \sim N^{\nu_\perp/\nu_\parallel} \Phi\left({t\over N}\right), \ \
W \sim N^{\nu_\perp/\nu_\parallel} \Psi\left({t\over N}\right),
\label{eq:HWdp}
\end{align}
where the scaling function behave as 
\begin{equation}
\Phi(x)\sim \Psi(x) \sim \left\{
\begin{array}{ll}
x^{\nu_\perp/\nu_\parallel} & \ {\rm for} \ x\ll 1,\\
{\rm const.} & \ {\rm for} \ x\gg 1.
\end{array}
\right.
\label{eq:Phidp}
\end{equation}
Note that the scaling exponent $\nu_\perp/\nu_\parallel \simeq 0.633$ and the dynamic exponent $z$ is $1$ in this argument~\cite{PhysRevE.51.4655}.

The scaling behavior of $H$ and $W$ in the lattice imbibition model with tension ($I>0$) is similar to Eqs.~(\ref{eq:HWdp}) and (\ref{eq:Phidp}) in the intermediate regime: Both the scaling exponent $B_*$ and $\beta_*$ are close to the value of $\nu_\perp/\nu_\parallel\simeq 0.633$, which indicates that our imbibition model with tension might be in the same universality class as the DP clusters. However, in contrast to interface growth in a disordered environment (DPD) the pinning in our model is not static but dynamically generated and the driving force decreases with time as the pressure gradient decreases. Moreover, the dynamic exponents $z_H$ and $z_W$ deviate from $1$ for the case of strong disorder ($\Delta=0.4$) of the pore radii considered in this work. 

\section{Conclusion and discussion}
\label{sec:discussion}

In this work we introduced a lattice model for spontaneous imbibition, which follows the same design principles as other interface growth models on a lattice. The emergent similarities facilitate a comparison of the characteristics of imbibition front propagation with other growth models in random environments and help to identify a minimal set of physical mechanisms that could explain the experimental observations~\cite{gruener12}. In particular, we have shown quantitatively how the disorder in the pore thickness and the  lateral correlation of hydrostatic pressure caused by liquid-volume conservation affect the roughening of the imbibition front in the elongated-pore systems. Also we found that the presence of an effective tension changes the scaling properties of  the front and leads to multiple crossovers in the dynamical evolution of height and roughness. The effective tension suppresses the formation of voids and smoothes the clusters of filled pores. As a result, with tension the imbibition front propagates faster and the roughness increases faster in the intermediate-time regime than that without the effective tension. Ultimately the imbibition front stops to propagate. By extensive simulations, we have identified the scaling behavior depending on the strength of the tension and the system size. 

Despite the approximation adopted in computing the hydrostatic pressure, the simulation is still time-consuming, which restricted the study to small system sizes. As a result it is hard to increase the accuracy of  the estimated scaling exponents. While some scaling exponents in this study are close to those of the directed percolation class, there are differences as well: the pinning centers are not quenched but dynamically generated and the dynamic exponent is found to deviate from that of the DP class for strong disorder in the pore radii, suggesting a new universality class. Thus, it remains an open question whether  the proposed lattice imbibition model belongs to the directed percolation universality class or represents a novel one. 

Since our model predicts that an effective tension changes the asymptotic behavior of the front propagation drastically it would be interesting to study experimentally the spontaneous imbibition of nano-porous materials in which one could change the aspect ratio of pores (i.e. the typical ratio between pore radius and pore length): The Vycor glass studied in \cite{gruener12} contained a network of elongated pores (aspect ratio around 5) and one would expect an effective tension to be absent since menisci in different pores are well separated. For aspect ratios around 1 the size of the menisci become comparable to the length of the pores (depending on wetting angle) and the menisci at junctions start to coalesce leading to a connected liquid-air interface. We propose that our imbibition model captures this situation by a non-vanishing effective tension - and predict that in this case neither the Lucas-Washburn law nor the roughening exponent $\beta=1/2$ continues to hold. With nano-porous materials one should even be able to reach the predicted asymptotic late-time regime, for which our model predicts the imbibition front propagation to cease.

\acknowledgements
This work was supported by the National Research Foundation of Korea (NFR) grants funded by the Korean Government (MEST) (No. 2012R1A1A2005252 (DSL)).

\vfill
\eject

\bibliography{imbibition4}

\end{document}